\documentclass{wsi-jgmmp}
\usepackage{bbm}
\usepackage{pdfsync}
\usepackage{ulem}
\usepackage{float}
\usepackage{color}
\usepackage{xcolor}
\usepackage[breaklinks,colorlinks,backref]{hyperref}
\hypersetup{
   colorlinks, 
   linktoc=all, 
   linkcolor=red,  
 citecolor=hyptxt,
 urlcolor=blue}
 \hypersetup{
 citebordercolor=,
 filebordercolor=green,
 linkbordercolor=blue
}
\definecolor{hyptxt}{rgb}{0.7, 0.4, 0.9}
\usepackage{bibtopic}
\usepackage{cite}
\DeclareFontFamily{U}{mathx}{}
\DeclareFontShape{U}{mathx}{m}{n}{<-> mathx10}{}
\DeclareSymbolFont{mathx}{U}{mathx}{m}{n}
\DeclareMathAccent{\widehat}{0}{mathx}{"70}
\DeclareMathAccent{\widecheck}{0}{mathx}{"71}

\newcommand{\beprop}{\begin{prop}}
\newcommand{\enprop}{\end{prop}}
\newcommand{\befor}{\begin{form}}
\newcommand{\enfor}{\end{form}}
\newcommand{\bprf}{\begin{proof}}
\newcommand{\eprf}{\end{proof}}
\newcommand{\ii}{\mathsf{i}}

\def\R{\mathbb{R}}
\def\N{\mathbb{N}}
\def\C{\mathbb{C}}
\def\Z {\mathbb{Z}}

\def\lg{\langle }
\def\rg{\rangle }

\def\vap{\mathfrak{w}}

\def\ud{\mathrm{d}}

\def\sfP{\mathsf{P}}
\def\sfMv{\mathsf{M}^{\vap}}
\def\sfMvP{\mathsf{M}^{\vap_\mathsf{P}}}
\def\sfMvcs{\mathsf{M}^{\vap_{\phi}}}

\def\bu{\mathbbm{1}}

\numberwithin{equation}{section}
\begin{document}

\markboth{Murenzi, Zlotak, Gazeau }
{Quantum discrete Torus}

%
\catchline{}{}{}{}{}
%

\title{WEYL-HEISENBERG COVARIANT QUANTIZATION FOR THE DISCRETE TORUS}

\author{ROMAIN MURENZI}

\address{Dep. Physics, Worcester Polytechnic Institute,\\ Worcester, MA 01609, USA\\
\email{rmurenzi@wpi.edu} }

\author{AIDAN ZLOTAK}

\address{Dep. Physics, Worcester Polytechnic Institute,\\ Worcester, MA 01609,USA\\
\email{ahzlotak@wpi.edu}}
\author{JEAN-PIERRE GAZEAU}

\address{ Astroparticule et Cosmologie, CNRS, Universit\'e Paris Cit\'e, \\ Paris, F-75013, France\\
\email{gazeau@apc.in2p3.fr}}
\maketitle

\begin{history}
\received{(Day Month Year)}
\revised{(Day Month Year)}
\end{history}

\begin{abstract}{Covariant  integral quantization is implemented for  systems whose phase space is  $\Z_{d}\times\,\Z_{d}$, i.e., for systems moving on the discrete periodic set $\Z_d= \{0,1,\dotsc d-1$ mod $d\}$.  The symmetry  group of this phase space is the periodic discrete version of  the Weyl-Heisenberg group, namely  the central extension of the abelian group $\Z_d\times\,\Z_d$. In this regard, the phase space is viewed as the left coset of the group with its center. The non-trivial unitary irreducible representation of this group, as acting on $L^2(\Z_{N})$,  is square integrable on the phase phase.  We  derive the  corresponding covariant  integral quantizations from (weight) functions  on the phase space, and display their phase space portrait.
}\end{abstract}



\tableofcontents

\section{Introduction}
\label{intro}
In this contribution, we explore quantum models for the unit-scale discrete torus $Z_{d} \times Z_{d}$,
which serves as a finite phase space ideally suited to systems with a finite number of states \cite{wootters, Gross1, Gross2, Zak}. This structure is highly relevant to quantum systems such as spin systems or those described by finite-dimensional Hilbert spaces. The discrete torus is particularly useful for defining discrete Wigner functions, which allow for the representation of quantum states in phase space \cite{wootters,gross}. This discrete approach is critical for analyzing quantum states in contexts where continuous phase space methods do not apply \cite{Vourdas1,Vourdas2,Vourdas3, Zak, Albert1}. The discrete torus also provides a valuable simplified model for studying quantum chaos and quantum maps, such as the quantum cat map \cite{keating}. These models can exhibit complex dynamical behaviors similar to classical chaotic systems, making them a key tool in exploring the quantum-classical correspondence. The discrete torus plays an important role in investigating the relationship between quantum and classical systems.

In quantum information and computation, the discrete torus is pivotal for studying quantum error correction, quantum algorithms, and quantum codes \cite{nielsen1,gottesman1, Kitaev}. By examining the evolution and transformation of quantum states in this discrete setting, researchers can develop efficient computational techniques. Additionally, the discrete structure of the torus is closely connected to modular arithmetic, which is a foundational element in quantum algorithms like Shor’s algorithm \cite{Shor}, which is widely used in cryptography. Mathematically, the discrete torus has deep connections with the representation theory of finite quantum groups. This relationship facilitates the study of symmetries and invariants in quantum systems, contributing to a more profound understanding of quantum symmetries.

This work employs the Weyl-Heisenberg covariant integral quantization method, with a particular focus on the use of Perelomov coherent states\cite{Perelomov}. Covariant integral quantization provides a systematic way to map classical observables to quantum operators using coherent states while preserving the symmetries of the underlying phase space. Coherent states act as a bridge between classical and quantum mechanics, offering a representation of phase space that retains much of the classical intuition \cite{Ali,Gazeau_coherent}. For the discrete torus, the construction of coherent states using the discrete Weyl-Heisenberg group offers a way to explore quantum states in this finite phase space. The discrete Weyl-Heisenberg group provides a group-theoretical framework for the quantization of the torus, allowing for unitary representations that correspond to translations in discrete phase space \cite{wootters}. This group structure is essential for defining discrete analogs of continuous transformations in quantum mechanics and for preserving the algebraic structure of phase space, particularly the commutation relations.

Coherent states on the discrete torus enable the construction of discrete probability distributions, such as the discrete Husimi functions, as well as quasi-probability distributions like discrete Wigner functions. In the domain of quantum information processing, coherent states and their associated quantization methods play a central role in encoding, transmitting, and manipulating quantum information within finite-dimensional Hilbert spaces. The discrete Weyl-Heisenberg group is also instrumental in designing quantum error correction techniques and analyzing noise processes in quantum channels. This approach is similarly important in signal processing, where the discrete torus and its quantization methods are applied to analyze and transform discrete signals.

Coherent states on the discrete torus can be considered finite-dimensional analogs of the coherent states of the quantum harmonic oscillator\cite{Werner, Richard_AU, Folland}. This analogy allows for the study of harmonic oscillator dynamics in a discrete context, which is particularly useful when modeling systems with periodic or finite boundary conditions. In the context of spin systems, the discrete Weyl-Heisenberg group provides a natural setting for defining related ``spin coherent states''. These states help in analyzing spin dynamics within quantum spin networks, which are essential models in both condensed matter physics and quantum computing \cite{Kitaev}.


We have organized the discussion of the discrete torus as follows:  Section \ref{survey} gives a review of covariant integral quantization and how it applies to the discrete torus. In section \ref{weyl} we consider the Weyl displacement operator and the action it takes on $L^2(\Z_d)$. Next in section \ref{quant} the details of how covariant integral quantization applies to the discrete torus are discussed. Section \ref{portrait} details how the semi-classical portrait of phase space operators can be found and why they are of interest. Finally,  with section \ref{weight} we illustrate the discussed formalism with two examples, namely the so-called coherent state weight and the weight distribution related to the parity operator.

\section{(Covariant) integral quantization: a survey}
\label{survey}

In this section we give a short review  of what we understand by covariant integral quantization in view of the application to the set $\Z_{d}\times\,\Z_{d}$.  

Let $(X,\mu)$ be a measure space and $\mathcal{H}$ be a (separable) Hilbert space.   An operator-valued function
\begin{equation}
\label{map1}
X\ni x \mapsto \mathsf{M}(x)\ \mbox{acting in} \ \mathcal{H}\, ,
\end{equation}
resolves the identity operator $\mathbbm{1}$ in  $\mathcal{H}$ with respect the measure $\mu$ if
\begin{equation}
\label{resUn}
\int_X \mathsf{M}(x)\, \mathrm{d} \mu(x)= \mathbbm{1}
\end{equation}
holds in a weak sense.

Integral quantization based on \eqref{map1} is   the linear map of a function on $X$ to an operator in $\mathcal{H}$ which is defined by
\begin{equation}
\label{intqgen}
f(x) \mapsto \int_X f(x) \mathsf{M}(x)\, \mathrm{d} \mu(x):= A_f\, , \quad 1 \mapsto \mathbbm{1}\,.
\end{equation}
If the operators $\mathsf{M}(x)$ in \ref{resUn}
are nonnegative and bounded,  one says that  they form a (normalised) positive operator-valued measure (POVM) on $X$.  If they are further
unit trace-class  for all $x\in X$, i.e., if the $\mathsf{M}(x)$'s are density operators, then the map
\begin{equation}
\label{sempor}
f(x) \mapsto  \check{f}(x):= \mathrm{tr}(\mathsf{M}(x)A_f) = \int_X f(x^{\prime})\,\mathrm{tr}(\mathsf{M}(x)\mathsf{M}(x^{\prime}))\, \mathrm{d} \mu(x^{\prime})
\end{equation}
is a local averaging of the original $f(x)$ (which can very singular, like a Dirac) with respect to the probability distribution on $X$,
\begin{equation}
x^{\prime} \mapsto \mathrm{tr}(\mathsf{M}(x)\mathsf{M}(x^{\prime}))\,.
\end{equation}
This averaging, or semi-classical portrait of the operator $A_f$,  is in general a regularisation, depending of course on the topological nature of the measure space $(X,\mu)$ and the functional properties of the $\mathsf{M}(x)$'s.

Let us now assume  that $X=G$ is a Lie group with left Haar measure ${\rm d}\mu(g)$, and let $g \mapsto U_g$ be a unitary irreducible representation (UIR) of $G$ in a Hilbert space $\mathcal{H}$. Let $\mathrm{M}$ be a bounded self-adjoint operator on $\mathcal{H}$ and let us define $g$-translations of $M$ as
\begin{equation}
\label{eqMg}
\mathrm{M}(g)= U_g \mathrm{M} U_g^\dagger\,.
\end{equation}
Suppose that the  operator
\begin{equation}
\label{intgrR}
R:= \int_G  \, \mathrm{M}(g)\,\ud\mu(g) \, ,
\end{equation}
is defined in a weak sense. From the left invariance of $\ud\mu(g)$  the operator $R$ commutes with all operators $U(g)$, $g\in G$, and so, from Schur's Lemma, we have the resolution of the unity up to a constant,
\begin{equation}
\label{resunitG}
R= c_{ \mathrm{M}}\mathbbm{1} \,.
\end{equation}
The constant $c_{ \mathrm{M}}$ can be found from the formula
\begin{equation}
\label{calcrho}
c_{ \mathrm{M}} = \int_G  \, \mathrm{tr}\left(\rho_0\, \mathrm{M}(g)\right)\, \ud\mu(g)\, ,
\end{equation}
where $\rho_0$ is a given unit trace positive operator. $\rho_0$ is  chosen, if manageable,  in order to make the integral convergent.  Of course, it is possible that no such finite constant exists for a given $\mathrm{M}$, or worse, it can not exist for any $\mathrm{M}$ (which is not the case for square integrable representations).
Now, if $c_{ \mathrm{M}}$ is finite and positive, the true resolution of the identity follows:
\begin{equation}
\label{Resunityrho}
\int_G \, \mathrm{M}(g) \,\ud \nu(g) = \mathbbm{1}\,, \quad \ud \nu(g):= \ud\mu(g)/c_{ \mathrm{M}}\, .
\end{equation}
For instance, in the case of a square-integrable unitary irreducible representation $U: g \mapsto U_g$, let us pick a unit vector $| \psi \rangle$ for which $c_{\mathrm{M}} = \int_G {\rm d}\mu(g) |\langle \psi | U_g \psi \rangle |^2 < \infty$, i.e $| \psi \rangle$ is an admissible unit vector for $U$. With $\mathrm{M} = |\psi \rangle \langle \psi |$ the resolution of the identity (\ref{Resunityrho}) provided by the family of states
$| \psi_g \rangle = U_g | \psi \rangle$ reads
\begin{equation}
\label{resunpsipsi}
\int_G |\psi_g \rangle \langle \psi_g | \frac{{\rm d} \mu(g)}{c_{\mathrm{M}}} =\mathbbm{1}   \,.
\end{equation}
Vectors $| \psi_g \rangle$ are named (generalized) coherent states (or wavelet) for the group $G$. \\
The equation (\ref{Resunityrho}) provides an integral quantization of complex-valued functions on the group $G$ as follows
\begin{equation}
\label{quantiz}
f \mapsto A_f = \int_G \mathrm{M}(g) f(g) \frac{{\rm d} \mu(g)}{c_{\mathrm{M}}} \,.
\end{equation}
Furthermore, this quantization is \emph{covariant} in the sense that:
\begin{equation}
\label{covar}
U_g A_f U_g^\dagger = A_F \quad \text{where} \quad F(g') = (U_g f)(g') = f(g^{-1} g')\,,
\end{equation}
 i.e. $U_g: f \mapsto F$ is the regular representation if $f \in L^2(G, {\rm d}\mu(g))$.

Concerning the group $\mathrm{H(\Z_{d})}$ we are considering in the present paper, an adaption of this above material   is necessary  in the sense that we have to replace with its coset $\Gamma_d:=\Z_{d}\times \Z_{d}$, which amounts to  replace its UIR with its projective version.

\section{Overview:  Weyl (displacement) operator acting on \texorpdfstring{\(L^2(\Z_{d})\)}{L2(Zd)}}
\label{weyl}

Following Vourdas, Gazeau, and Feichtinger,\cite{Vourdas1,Vourdas2,Vourdas3} we consider a  system (resp. a signal) where position (resp. time) and momentum (resp. frequency)  take values in the ring $\Z_d=\Z/d\Z=\{0,1,...,d-1\}$ of the integer modulo $d$.
The Hilbert  space $\mathcal{H}\equiv L^2(\Z_d)$ of the quantum states of this  system is the vector space of  $d$-periodic complex sequences
\begin{equation}
\label{Cseq}
\phi= (\phi(1),\phi(2), \dotsc, \phi(d))\, , \quad \phi(l)\in \C\, , \quad \phi(l+d)=\phi(l)\, , 
\end{equation} 
equipped with the scalar product
\begin{equation}
\label{inprod}
\lg\phi_1|\phi_2\rg=\sum_{l=0}^{d-1}\overline{\phi_1(l)}\phi_2(l)=\sum_{l\in\Z_d}\overline{\phi_1(l)}\phi_2(l)\,. 
\end{equation}
Note that the Hilbertian Dirac bracket formalism  and notations are   occasionally used in this paper: 
\begin{equation}
\label{dibracket}
\phi(l) \equiv \lg l | \phi\rg\, ,   \quad  \lg l|l^{\prime}\rg = \delta_{ll^{\prime}}\, , \quad \sum_{l=0}^{d-1} |l\rg\lg l |= \bu\,.
\end{equation}

An orthonormal basis of Fourier basis of $\mathcal{H}$  is provided by the $d$ Fourier exponentials $\left(e^{d}_{k}\right)_{0\leq k\leq d-1}$:
\begin{equation}
\label{fourd}
e^{d}_{k}(l)=\frac{1}{\sqrt{d}}e^{\ii\frac{\,2\pi}{d}\,k\,l}\, ,
\end{equation}
with
\begin{equation}
\label{orthog}
\langle\,e^{d}_{k}|\,e^{d}_{k^{\prime}}\rangle=\frac{1}{d} \sum_{l=0}^{d-1}e^{\ii\,\frac{2\pi}{d}(k^{\prime}-k)l}=\frac{1}{d}\frac{1-e^{\ii\,2\pi(k^{\prime}-k)}}{1-e^{\ii\,\frac{2\pi}{d}(k^{\prime}-k)}}=\delta_{k\,k^{\prime}}\,,
\end{equation}
and 
\begin{equation}
\label{orthoFour}
\sum_{k=0}^{d-1} |e^{d}_{k}\rg\lg e^{d}_{k}| = \bu\, . 
\end{equation}
The Fourier transform of $\phi=(\phi(l))_{l\in\Z_d}$ and its inverse are defined as:
\begin{align}
\label{fourt}
&(\mathcal{F}\phi)(k)=\hat{\psi}(k)=\langle\,e^{d}_{k}|\phi\rangle= \frac{1}{\sqrt{d}}\sum_{l=0}^{d-1}e^{-\ii\,\frac{2\pi}{d}\,k\,l}\phi(l)\, , \\\label{fourtI}
&(\mathcal{F}^{-1}\hat\phi)(k)=\phi(k)=\langle\,\overline{e^{d}_{k}}|\hat\phi\rangle= \frac{1}{\sqrt{d}}\sum_{l=0}^{d-1}e^{\ii\,\frac{2\pi}{d}\,k\,l}\hat\phi(l)\,. 
\end{align}
The Fourier transform of the Fourier basis elements is given by:
\begin{equation}
\label{fourfour}
(\mathcal{F}e^{d}_{k_0})(k)= \hat{e}^{d}_{k_0}(k)= \langle\,e^{d}_{k}|\,e^{d}_{k_0}\rangle= \delta_{k\,k_0}\equiv \lg k | k_0\rg\, . 
\end{equation}
Hence, the dual basis  to the Fourier basis is precisely the set of kets $\{\widehat{e^{d}_{k}}\equiv |k\rg\, , \, k\in \Z_d\}$, defined as the discrete functions $\hat{e}^{d}_{k}(l)\equiv \lg l | k\rg=\delta_{kl}$. Reciprocally, 
\begin{equation}
\label{fourk}
(\mathcal{F}^{-1} \hat{e}^{d}_{k})(l) = e^{d}_{k}(l)\, . 
\end{equation}
These formulas are crucial for understanding  the definition and properties given in Eq. \eqref{dibracket} and the nature of  the Cartesian product $\Gamma_d= \Z_d\times\Z_d$ viewed as a phase space.

It is easy to show that the symmetry for this system is the discrete Weyl-Heisenberg group $\mathrm{H}(\Z_{d})$. The latter is viewed as  the set of triplets $(s, m,n)\in \R\times\Z_{d}\times \Z_{d}$ equipped with the internal law:
\begin{equation}
\label{intlaw}
(s,m,n)\,(s^{\prime}, m^{\prime},n^{\prime})= \left(s+s^{\prime} +\frac{m\,n^{\prime}-m^{\prime}n}{2},m+m^{\prime}\,\mathrm{mod} \, d,n+n^{\prime} \, \mathrm{mod}\,d\right)\,.
\end{equation}
For the sake of simplicity, we will omit the ``mod $d$'' notation from this point onward.

The neutral element is $e=(0,0,0)$ and the inverse is given by $$(s,m,\theta)^{-1}= (-s,-m\, \mathrm{mod}\, d,-n\,  \mathrm{mod}\, d)\,. $$
The Abelian one-parameter subgroup $C= \{(s, 0,0)\, , \, s\in \R\}$ is the center of $\mathrm{H}(\Z_{d})$.  

Besides the trivial one, the group $\mathrm{H}(\Z_{d})$ admets  the following  unitary  irreducible representation   acting on $\mathcal{H}$ as:
\begin{align}
\begin{split}
\label{Operator_identities}
&(V(s,m,n)\psi)(l) =e^{\ii\,\frac{2\pi}{d}s}e^{-\ii\,\frac{\pi}{d}m\,n}e^{\ii\,\frac{2\pi}{d}m\,l}\psi(l-n)\,\,\\\nonumber 
&V(s,m,n)V(s^{\prime},m^{\prime},n^{\prime})= V(s+s^{\prime}+\frac{m\,n^{\prime}-m^{\prime}n}{2},m+m^{\prime},n+n^{\prime})\,,\\
& V^{\dag}(s,m,n) = V(-s,-m,-n)= V((s,m,n)^{-1})\,.
\end{split}
\end{align}
As for the continuous case, this operator is the combination of two elementary operations, translation 
\begin{equation}
\label{transl}
(T_{l_0}\psi)(l)=\psi(l-l_0)\, , 
\end{equation}
and modulation (or translation in the Fourier space)
\begin{equation}
\label{mod}
(E_{k_0}\psi)(l)=e^{\ii \frac{2\pi}{d}k_0l}\psi(l) \ \Leftrightarrow \ \mathcal{F}(E_{k_0}\psi)(k) = \hat{\psi}(k-k_0)\, . 
\end{equation}
Note the commutation rule:
\begin{equation}
\label{comW}
T_{l_0}E_{k_0}= e^{-\ii\frac{2\pi}{d}k_0l_0}\, E_{k_0}T_{l_0}\,. 
\end{equation}
Due to the central parameter $s\in \R$,  the representation is not square integrable. But, it is square integrable on the phase space $\Gamma_d$. In this group  context, the latter is viewed as  the left coset $\Gamma_d=\mathrm{H}(\Z_{d})/C$ of $\mathrm{H}(\Z_d)$ with its center.
Hence, the representation  induces the  unitary ($\sim$ Weyl) operator $U$ defined on  $\Gamma_N$ by the section $s=0$ on the group, \textit{i.e.}, ${U}(m,n):= {V}(0,m,n)$:
\begin{equation}
\label{WeylD}
\begin{split}
(U(m,n)\psi)(l)&=e^{-\ii\,\frac{\pi}{d}m\,n}e^{\ii\,\frac{2\pi}{d}m\,l}\psi(l-n)\\
&= e^{-\ii\,\frac{\pi}{d}m\,n}\left(E_mT_n \psi\right)(l) = e^{\ii\,\frac{\pi}{d}m\,n}\left(T_n E_m\psi\right)(l)\,,\\
U^{\dag}(m,n)&= U(-m,-n)\,.
\end{split}
\end{equation}
Let us now give three useful formulae for the sequel.
Firstly the composition formula, where the symplectic underlying structure of the WH symmetry is put into evidence in the phase factor:
\begin{equation}
\label{UUp}
\begin{split}
U(m,n)\,U(m^{\prime},n^{\prime})&= e^{\ii\frac{\pi}{d}(mn^{\prime}-nm^{\prime})}\,U(m+m^{\prime},n+n^{\prime})\\
\end{split}
\end{equation}
\begin{equation}
\label{UUUp}
{U}(m^{\prime},n^{\prime}){U}(m,n)\,{U}^{\dag}(m^{\prime},n^{\prime})= e^{-\ii\frac{2\pi}{d}(mn^{\prime}-m^{\prime}n)}U(m,n)\,.
\end{equation}
\begin{equation}
\label{TrU}
\mathrm{Tr}[U(m,n)]=d\,\delta_{m\,0}\delta_{n\,0}\,.
\end{equation}
Applying $U(m,n)$  to the Fourier basis elements and their reciprocal yields: 
\begin{align}
\label{Ufour}
(U(m,n)e^{d}_{k})(l)&= 
e^{\ii\,\frac{\pi}{d}m\,n} \,e^{d}_{k+m}(l-n)\, , \\ 
\label{Udir}(U(m,n)\hat{e}^{d}_{k})(l)&= e^{\ii\,\frac{\pi}{d}m(2k+n)} \delta_{l k+n}=  e^{\ii\,\frac{\pi}{d}m(2k+n)} \hat{e}^{d}_{k+n}(l)\, . 
\end{align}
One then derives the matrix elements of the displacement operator $U(m,n)$ in the respective  Fourier and reciprocal  bases:
\begin{align}
\left(U(m,n)\right)_{kk^{\prime}}&= \lg e^{d}_{k}| U(m,n)e^{d}_{k^{\prime}}\rg= e^{\ii\,\frac{\pi}{d}mn}e^{-\ii\,\frac{2\pi}{d}kn} \delta_{k, k^{\prime}+m}\,,\\
\label{matelUD}
\left(\widehat{U(m,n)}\right)_{kk^{\prime}}&= \lg \hat{e}^{d}_{k}| U(m,n)\hat{e}^{d}_{k^{\prime}}\rg= e^{\ii\,\frac{\pi}{d}m(k+k^{\prime})}\delta_{k, k^{\prime}+n}\,.
\end{align}
For d=2, we get the Pauli matrices, including the unit matrix:
\begin{align}
&\left(U(0,0)\right)=I_{2},  
\left(U(0,1)\right)=
\begin{pmatrix}
1 & 0  \\
0 & -1
\end{pmatrix}=\sigma_3\,, \\\nonumber
&\left(U(1,0)\right)=
\begin{pmatrix}
0& 1 \\
1 & 0 
\end{pmatrix}\equiv \sigma_1,
\left(U(1,1)\right)=
\begin{pmatrix}
0 & i \\
-i & 0 
\end{pmatrix}\equiv \sigma_2\,
\end{align}

By direct algebraic manipulations  one proves that, for any unit-norm $\psi$ in $\mathcal{H}$, the family 
\begin{equation}
\label{psimn}
\{{U}(m,n)\psi\equiv \psi_{(m,n)}\, , \, (m,n)\in\Gamma_d\}
\end{equation}
constitutes an overcomplete  family, \textit{i.e.},  it
resolves the identity on $\mathcal{H}$  in the sense of \eqref{resUn} or \eqref{Resunityrho} or \eqref{resunpsipsi}:
\begin{equation}
\label{resunZS}
\frac{1}{d}\sum_{(m,n)\in\Gamma_d}\, |\psi_{(m,n)}\rangle\langle \psi_{(m,n)}| = \bu\,, \quad \Vert \psi \Vert = 1\, .
\end{equation}
This fundamental property can be extended  to the $U$-transported versions of a density operator $\rho = \sum_i p_i |\psi_i\rangle\langle\psi_i|$ ($\sim$ mixed quantum state) on $\mathcal{H}$, where $\mathcal{J}\ni i\mapsto p_i$ is a discrete probability distribution on an index set $\mathcal{J}$ (one could as well deal with a continuous distribution), and all the states $|\psi_i\rangle\in \mathcal{H}$ are unit norm: 
\begin{equation}
\label{densop}
\frac{1}{d}\sum_{(m,n)\in\Gamma_d}\, \rho(m,n)= \bu\, , \quad  \rho(m,n):={U}(m,n)\rho{U}^{\dag}(m,n) \,. 
\end{equation}
This implies that for any $\phi$, $\psi$ in $\mathcal{H}$ we have:
\begin{equation}
\label{resunZSpp}
\frac{1}{d}\sum_{(m,n)\in\Gamma_d}\,\,\vert\lg\phi | \rho(m,n)|\phi^{\prime}\rangle\vert^2= \Vert\phi\Vert^2 \Vert\phi^{\prime}\Vert^2\,.
\end{equation}
In \eqref{resunZS} the state  $\psi_{(m,n)}$ is a coherent state (CS) for the group $H(\Z_d)$ in the sense given by  Perelomov \cite{Perelomov}, and the picked unit norm   $\psi$ is  called a fiducial vector.
The projection  of $\phi\in \mathcal{H}$  on $\psi_{m,n}$, namely
\begin{equation}
\label{pspphi}
\langle\,\psi_{(m,n)}|\phi\rangle
= \sum_{l=0}^{d-1}\, \overline{U(m,n)\psi(l)}\phi(l)
=  \sum_{l=0}^{d-1}\,e^{\ii\,\frac{\pi}{d}m\,n}e^{-\ii\,\frac{2\pi}{d}m\,l}\overline{\psi(l-n)}\phi(l)\, ,
\end{equation}
is the  phase space representation of  $\phi$ with respect to the family of  coherent states $\psi_{(m,n)}$, and $\vert\langle\,\psi_{m,n}|\phi\rangle\vert^2/d$ is the corresponding Husimi function.  More generally, $\frac{1}{d}\mathrm{Tr}\left(|\phi\rg\lg \phi| \rho(m,n)\right)= \frac{1}{d}\lg\phi|\rho(m,n)|\phi\rg$ is the Husimi function of  $\phi$ in its phase-space representation provided by the family $\{\rho(m,n)\}_{(m,n)\in \Gamma_d}$. 
In the sequel we restrict our analysis by using CS's only. The extension of our results to  transported density operators is straightforward. 
 
 We now summarize the most important features of the CS analysis of the elements of $\mathcal{H}$ resulting from the resolution of the identity \eqref{resunZS}.
\begin{proposition}
\label{main_properties}
For any elements  $\phi, \psi\in\,\mathcal{H}$, such that $\vert\vert\psi\vert\vert=1$, the map
\begin{equation}
\phi\rightarrow\langle\psi_{(m,n)}\vert\phi\rangle \equiv \Phi(m,n)
\end{equation} satisfies the following properties:
\begin{itemize}
\item it is an isometry:
\begin{align}
\label{isom}
\vert\vert\phi\vert\vert^2=\sum_{l=0}^{d-1}\vert\phi(l)\vert^2= \frac{1}{d}\sum_{(m,n)\in\Gamma_d}\vert{\langle\psi_{(m,n)}|\phi\rangle\vert}^2\,,
\end{align}
\item it can be inverted on its range:
\begin{equation}
\label{invert}
\phi(l)= \frac{1}{d}\sum_{(m,n)\in\Z_d\times\Z_d}\,\langle\psi_{(m,n)}|\phi\rangle \,\psi_{(m,n)}(l)\, ,
\end{equation}
\item its range lies in the  reproducing kernel space characterised by:
\begin{equation}
\label{reprod}
\langle\psi_{(m,n)}|\phi\rangle= \frac{1}{d}\sum_{(m,n)\in \Gamma_d}\,
K_{\psi}((m,n), (m^{\prime},n^{\prime}))\Phi(m^{\prime},n^{\prime})\,,
\end{equation}
\end{itemize}
with  kernel $K_{\psi}((m,n), (m^{\prime},n^{\prime}))=\langle\psi_{(m,n)}|\psi_{(m^{\prime},n^{\prime})}\rangle$. 
\end{proposition}
Note  the expression for the kernel:
\begin{equation}
\label{Kpsi}
K_{\psi}((m,n), (m^{\prime},n^{\prime}))=A(m,m^{\prime},n,n^{\prime})\Psi(m-m^{\prime},n-n^{\prime})\, ,
\end{equation}
where
\begin{equation}
\label{Amm}
A(m,m^{\prime},n,n^{\prime})= e^{\ii\,\frac{\pi}{d}(m(n-n^{\prime})-(m-m^{\prime})n^{\prime})}= = e^{\ii\,\frac{\pi}{d}((m-m^{\prime})(n-n^{\prime})-(mn^{\prime}) -nm^{\prime})}\, , 
\end{equation}
and
\begin{equation}
\label{Phimn}
\Psi(m-m^{\prime},n-n^{\prime})= \sum_{l\in\Z_d}\,e^{-\ii\frac{2\pi}{d}(m-m^{\prime})l}\overline{\psi(l-(n-n^{\prime}))}\psi(l)\,. 
\end{equation}

Let us present some examples of fiducial vectors.
 \begin{itemize}
\item Constant function:
\begin{equation}
\phi(l)=\frac{1}{\sqrt{d}},
\end{equation}
and the reproducing kernel:
\begin{equation}
\left\lg\phi_{m\,n}\vert \phi_{m^{\prime} n^{\prime}}\right\rg=
\delta_{m\,m^{\prime}}e^{-\ii\frac{\pi}{d}m(n-n^{\prime})}.
\end{equation}
\item Kronecker vector:
\begin{equation}
\phi_{k_0}(l)=\delta_{l\,k_0}\,,
\end{equation}
and the reproducing kernel:
\begin{equation}
\langle\phi_{m\,n}\vert\phi_{m^{\prime}n^{\prime}}\rangle=
\delta_{n\,n^{\prime}}e^{-\ii\frac{2\pi}{d}\frac{(k_0+n)}{2}(m-m^{\prime})}.
\end{equation}
\item Discrete wave plane vector:
\begin{equation}
\phi_{k_0}(l)= \frac{e^{\ii\,\frac{2\pi}{d}k_0\,l}}{\sqrt{d}}.
\end{equation}
with its reproducing kernel
\begin{align}
    \left\lg\phi_{m\,n}\vert \phi_{m^{\prime}n^{\prime}}\right\rg = \delta_{m\,m^{\prime}}e^{-\ii\frac{\pi}{d}(n'-n)(m+2k_0)}.
\end{align}
\item Periodization of $g\in l^2(\Z)$:
\begin{equation}
\phi(l)= \sum_{m=-\infty}^{+\infty}g(l+m\,d).
\end{equation}

\item\, Discrete periodized  ($L^2$ normalized ) Gaussian
\begin{equation}
\phi(l) = g_{\kappa} (l) = \frac{1}{\sqrt{\vartheta_3 \left(0,\frac{2i}{\kappa d}\right)}} \sum_{n = - \infty}^{\infty} e^{\ii \frac{2\pi}{d}nl}e^{-\frac{\pi}{\kappa d}n^2} ,
\end{equation}
with reproducing kernel
\begin{align}
    &\left\lg\phi_{m\,n}\vert \phi_{m^{\prime} n^{\prime}}\right\rg = \frac{d}{\sqrt{\vartheta_3 \left(0,\frac{2\ii}{\kappa d}\right)}} e^{\ii\frac{\pi}{d}(m\,n-m^{\prime})}e^{-\ii\frac{2\pi}{d}q^{\prime}(m^{\prime}-m)}e^{-\frac{\pi}{\kappa d}(m-m^{\prime})^2}\times \\\nonumber
&\vartheta_3\left(\frac{n-n^{\prime}}{d} - \frac{\ii(m-m^{\prime})}{\kappa d}, \frac{2\ii}{\kappa d} \right) ,\\\nonumber
\end{align}
where $\vartheta_{3}$ is the  third Jacobi theta function:
\begin{equation}
\vartheta_{3}(x,s)= \sum_{n\in\Z}\,e^{\ii\,2\pi\,n\,x}\,e^{\ii\pi\,s\,n^2}\,. 
\end{equation}
\item Discrete Dirichlet fiducial vector, $d=2j+1$:
\begin{align}
D_{(d,n)}(l)
= \frac{1}{\sqrt{d(2j+1)}}\sum_{m=-j}^{m=+j}\,e^{\ii\frac{2\pi}{d}\,m\,l}
=\frac{1}{\sqrt{d(2j+1)}}
\frac{\sin(\frac{\pi\,(2j+1)l}{d})}{\sin(\frac{\pi\,l}{d})},
\end{align}
and its reproducing kernel:
\begin{equation}
\left\lg\phi_{m\,n}\vert \phi_{m^{\prime} n^{\prime}}\right\rg=
\frac{d}{2j+1}\delta_{n\,n^{\prime}}e^{-\ii\frac{2\pi}{d}\frac{n(m-m^{\prime})}{2}}.
\end{equation}
\item Von Mises Fiducial Vector:
\begin{equation}
    \psi(l)=\frac{e^{\lambda\cos (\frac{2\pi}{d}l)}}{\sqrt{d\,I_{0}(2\lambda)}}\,.
    \label{VonMises} 
\end{equation}
\end{itemize}
\section{Covariant integral quantization on the discrete torus}
\label{quant}
\subsection{Quantization operators, Husimi and Wigner distributions}
Following previous works \cite{gazmur16,gazkoimur20,gazkoimur22}, we pick a function $\vap$, called weight (but not necessarily positive),  on the phase space $\Gamma$. We then define the  operator $\sfMv $, called ``quantization operator'', by
\begin{equation}
\label{affbop}
\sfMv =\frac{1}{d}\sum_{(m,n)\in\Z_d\times\Z_d}\,U(m,n)\vap(m,n)\, ,
\end{equation}
and we choose the weight such that the operator $\sfMv $ is bounded and symmetric, i.e., is self-adjoint on the Hilbert  space of physical states, which in our case $L^2(\Z_d)$. Given a weight $\vap$ and a function $f$:
\begin{equation}
 f: (p,q)\in\Gamma\rightarrow \C
\end{equation}
 defined on the phase space, one would like to define an operator defined on $L^2(\Z_d)$:
\begin{equation}
 A_f^{\vap}:\phi\rightarrow A^{\vap}_f\psi.
\end{equation}
Question: which combinations of $\vap$ and $f$ give rise to:
\begin{itemize}
\item Auto-adjointness,
\begin{equation}
 {A_f}^{\vap}= \left({A_f}\right)^{\vap{^\dag}}
\end{equation}
\item Density operators (Quantum states),
\begin{equation}
 A_f^{\vap}\geq 0\, , \quad \mathrm{Tr}\left[A_f^{\vap}\right]=1
\end{equation}
\end{itemize}
 
\begin{proposition}
\begin{itemize}
\item With the assumption  that the weight $\vap$ has been chosen such that the operator $\sfMv$, defined by \eqref{affbop}, is bounded:
The  operator $\sfMv$ is the ``integral" operator:
\begin{equation}
(\sfMv \psi)(m)=\sum_{m^{\prime}}\,\mathcal{M}^{\vap}(m,m^{\prime})\psi(m^{\prime})\, ,
\end{equation}
where the kernel ${\mathcal M}^{\vap}(m,m^{\prime})$ is given by:
\begin{equation}
{\mathcal M}^{\vap}(m, m^{\prime})= \frac{1}{d}\sum_{n\in\Z_d}\,\vap(n,m-m^{\prime})\,e^{\ii\frac{\pi}{d}(n(m+m^{\prime}))}
\equiv\frac{1}{\sqrt{d}}\widetilde{\vap}_{p_{1}}\left(\frac{m+m^{\prime}}{2},m-m^{\prime}\right)\, .
\end{equation}
Here, $\widetilde{\vap}_{p_{1}}$ is the  inverse discrete Fourier transform of $\vap$ with respect to the first variable.
\item  The operator $\sfMv $ is symmetric if and only the weight satisfies:
\begin{align}
\sfMv ={\sfMv }^{\dag} \Leftrightarrow \overline{\vap(-m,n)}=\vap(m,-n)\,.
\end{align}
\item The operator $\sfMv $ is of unit trace, for $\vap(0,0)=1$:
\begin{equation}
\mathrm{Tr}(\sfMv )=\vap(0,0)=1\,.
\end{equation}
\end{itemize}
\end{proposition}
\bprf
\begin{enumerate}
\item[(i)]
The action of $\sfMv $ on $\psi$ is given by:
\begin{align*}
&(\sfMv \psi)(m)\\
&= \frac{1}{\sqrt{d}}\sum_{m^{\prime}\in\Z_d}\left(\frac{1}{\sqrt{d}}\sum_{n\in\Z_d}\,\vap(n,m-m^{\prime})e^{\ii\frac{\pi}{d}(n(m+m^{\prime}))}\right)\psi(m^{\prime})\\
&=\sum_{m^{\prime}\in\Z_d}{\mathcal M}^{\vap}(m,m^{\prime})\psi(m^{\prime})
\end{align*}
\item[(ii)]  The condition that $\sfMv $ be symmetric implies the following condition on the kernel:
\begin{equation*}
\overline{{\mathcal M}^{\vap}(m, m^{\prime})}= {\mathcal M}^{\vap}(m^{\prime},m)
\end{equation*}
which gives:
\begin{equation*}
\overline{\widetilde{\vap}_{p_{1}}\left(\frac{m+m^{\prime}}{2},m-m^{\prime}\right)}=\widetilde{\vap}_{p_{1}}\left(\frac{m+m^{\prime}}{2},-(m-m^{\prime}))\right).
\end{equation*}
\item[(iii)] Therefore the trace of $\sfMv $ corresponds to the integral of the kernel over its diagonal, that is:
\begin{align*}
&\sum_{m\in\Z_d} {\mathcal M}^{\vap}(m,m)= \frac{1}{d}\sum_{n\in\Z_d}\sum_{m\in\Z_d}\,\vap(m,0)e^{\ii\,\frac{2\pi}{d}mn}\\
&=\frac{1}{d}\sum_{m\in\Z_d}\,\vap(m,0)d\delta_{m\,0}=\vap(0,0).
\end{align*}
\end{enumerate}
\eprf
In turn, one retrieves the weight $\vap$ from the quantization operator $\sfMv$ through a tracing operation.
\begin{proposition}
\label{tretrw}
The  trace of the operator ${U}^{\dag}(m,n)\sfMv $ is given by:
\begin{equation}
\label{UMvptr}
\mathrm{Tr}[U^{\dag}(m,n)\sfMv ]=\vap(m,n)
\end{equation}
\end{proposition}
\bprf
To compute this trace, one uses the orthonormal basis $\left\{e^{d}_k(l)= \frac{1}{\sqrt{d}}e^{\ii\,\frac{2\pi}{d}kl}\right\}$ as follows
\begin{align*}
&\mathrm{Tr}[U^{\dag}(m,n)\sfMv ] =\sum_{k\in\Z_d}\langle\,e^d_{k}\vert\,U^{\dag}(m,n)\sfMv e^d_{k}\rangle
\end{align*}
\eprf
\begin{proposition}
The transported of $\sfMv$ by U, that is, $\sfMv(p,q)=U^{\dag}(p,q)\sfMv\, U(p,q)^{\dag}$, multiplied by $\frac{1}{d}$ resolves the identity:
\begin{equation}
\label{trs_sfMv}
A^{\vap}(m,n)= \frac{1}{d}\sfMv(p,q)\, \quad ;\sum_{(p,q)\in\Gamma}A^{\vap}(m,n) = \mathbbm{1}\, . 
\end{equation}
\end{proposition}

We give below two examples of weights and the related $\sfMv$ and $\sfMv(p,q)$.
\begin{definition}[Coherent state weight]
Given a coherent state $U(m,n)\phi$ with a fiducial vector $\phi$, the related reproducing kernel 
\begin{equation}
\vap_{\phi}(m,n)=\langle\,U(m,n)\phi|\phi\rangle
\end{equation}
 is called coherent state weight.
\end{definition}

\begin{proposition}[Husimi distribution]
For a coherent state weight $\vap_{\phi}$, the average of the corresponding of the operator $A^{\vap_{\phi}}(m,n)$ \eqref{trs_sfMv} on a quantum pure state $|\psi\rangle\langle\psi|$ is the Husimi distribution of the $|\phi\rangle$ projected on the coherent state $|U(m,n)\phi\langle$, that is:
\begin{equation}
\label{expl_weights}
\langle \psi|A^{\vap_{\phi}}(m,n)\psi\rangle=|\langle\,U(m,n)\phi|\psi\rangle|^2\,.
\end{equation}
\end{proposition}
Therefore the operator $\frac{1}{d}\sfMvcs$ and its transported $\frac{1}{d}\sfMvcs(p,q)$ by U, are density  operators.

\begin{definition}[Parity or Wigner weight]
The unit weight $\vap(m,n)=1$, is called the parity weight, or Wigner weight.
\end{definition}

In the following proposition, we will restrict to an odd dimension $d$.
\begin{proposition}[Wigner distribution]
For the parity weight $\vap_{\phi}(m,n)=1$, and an odd dimension d, the operator $\mathsf{P}$ is the parity operator $\mathsf{P}$ on $\Z_d$, and the average of the corresponding of the operator $A^{\vap_{\mathsf{P}}}(m,n)$ \eqref{trs_sfMv} on a quantum pure state $|\psi\rangle\langle\psi|$ is a Wigner distribution distribution on the phase space $W_{\psi}(m,n)$.

\begin{align}
&(\sfMvP\psi)=\psi(-l)\equiv (\mathsf{P}\psi)(l)\\\nonumber
W_{\psi}(m,n) &= \left\langle \psi|(\frac{1}{d}\sfMvP(m,n)\psi)\right\rangle\\\nonumber
&=\frac{1}{d}\sum_{l\in\Z_d}\,e^{\ii\frac{2\pi}{d}ml} \overline{\psi(n+\frac{l}{2})}\psi(n-\frac{l}{2})\,. 
\end{align}
\end{proposition}
This Wigner distribution is real and satisfies marginality that is:
\begin{equation}
\sum_{m\in\Z_d} W_{\psi}(m,n)=|\psi(n)|^2\,,\quad  \sum_{n\in\Z_s} W_{\psi}(m,n) =|\widehat{\psi(m)}|^2\,.
\end{equation}
\subsection{Covariant integral quantization from weight function}
\label{genform}
\subsubsection{General results}
\label{genqres}
We now  establish general formulas for the integral quantization issued from
a weight function $\vap(m ,n )$ on $\Gamma=\Z_d\times \Z_d$ yielding the self-adjoint operator $\sfMv$ defined in \eqref{affbop}.
\begin{definition}[Quantization map]
Given a weight $\vap$, the related quantization operator $\sfMv$, and its transported $\sfMv(m,n)$ by $U$,the corresponding ``quantization map" $A^{\vap}_f$ is defined by:
\begin{equation}
\label{genqvap}
f\mapsto A^{\vap}_f= \frac{1}{d}\sum_{(m,n)\in\Z_d\times\Z_d}\,f(m,n )\, \sfMv(m,n)\,.
\end{equation}
\end{definition}
We  have the following result.
\begin{proposition}
    
 $A^{\varpi}_f$ is the integral operator on $L^2(\Z_d)$
\begin{equation}
(A^{\vap}_f\psi)(l)=\sum_{m^{\prime},n^{\prime}}{\mathcal A}^{\varpi}(l, l^{\prime}) \psi(l^{\prime})\,, 
\end{equation}
and its  kernel is given by
\begin{align}
\label{kernel_A_f}
\mathcal{A}_{f}^{\varpi}(l,l^{\prime}) = \frac{1}{d} \sum_{m\in\Z_d}\vap(m,l-l^{\prime})\,\overline{F_{s}[f]}(m,l-l^{\prime}) e^{\ii\frac{\pi}{d}m(l+l^{\prime})}\,,
\end{align}
where $F_{s}$ is the symplectic discrete Fourier transform:
\begin{align}
\label{fourier_two_var}
f \mapsto F_s[f]\ , \hspace{10mm} F_s[f](m,n)= \frac{1}{d} \sum_{(m,n)\in\Z_d\times\Z_d} f(m^{\prime},n^{\prime})e^{-\ii \frac{2\pi}{d}(m^{\prime}n-mn^{\prime})}
\end{align}
One can also define the ``conjugate'' symplectic Fourier transform $\overline{F_s}[f](n,\varphi)$ as 
\begin{align}
    f \mapsto \overline{F_s}[f]\ , \hspace{10mm} \overline{F_s}[f](m,n) =\frac{1}{d} \sum_{(m,n)\in\Z_d\times\Z_d} f(m^{\prime},n^{\prime})e^{\ii \frac{2\pi}{d}(m^{\prime}n-mn^{\prime})}
\end{align}
It is important to note that $\overline{F_s}$ is not the inverse of $F_s$ since they are both their own inverse: $F_s^2 = \mathbbm{1}= \overline{F_s}^2$. 
\begin{align}
    \overline{F_s}F_s = F_s\overline{F_s} = \mathsf{P}_2\ , \hspace{10mm} \left(\mathsf{P}_2f\right)(m,n) = f(-m,-n)\,.
\end{align}

\end{proposition}
\bprf
Let us first write the ``integral operator" $A^{\vap}_f$ using the symplectic Fourier transform. We have
\begin{align*}
&A^{\vap}_f
= \frac{1}{d}\sum_{(m,n)\in\Z_d\Z_d} f(m,n )\, \sfMv(m,n)\\
&=\frac{1}{d^2} \sum_{(m^{\prime},n^{\prime})\in\Z_d\times\Z_d} \sum_{(m,n)\in\Z_d\times\Z_d} f(m,n)\vap(m^{\prime},n^{\prime})\, U(m,n)U(m^{\prime},n^{\prime})U^{\dag}(m,n)\\\nonumber
&=\frac{1}{d^2} \sum_{(m^{\prime},n^{\prime})\in\Z_d\times\Z_d} \sum_{(m,n)\in\Z_d\times\Z_d} f(m,n )\vap(m^{\prime},n^{\prime})\, e^{\ii\frac{2\pi}{d}(mn^{\prime}-m^{\prime}\,n)}U(m^{\prime},n^{\prime})\\\nonumber
&=\frac{1}{d} \sum_{(m^{\prime},n^{\prime})\in\Z_d\times\Z_d}\vap(m^{\prime},n^{\prime})\,[\frac{1}{d} \sum_{(m,n)\in\Z_d\times\Z_d} f(m,n ) e^{\ii\frac{2\pi}{d}(mn^{\prime}-m^{\prime}\,n)}]U(m^{\prime},n^{\prime})\\\nonumber
&=\frac{1}{d} \sum_{(m^{\prime},n^{\prime})\in\Z_d\times\Z_d}\vap(m^{\prime},n^{\prime})\,\overline{F_{s}[f]}(m^{\prime},n^{\prime}) U(m^{\prime},n^{\prime})\,. 
\end{align*}
Applying to $\psi$ gives:
\begin{align*}
&(A^{\vap}_f\psi)(l)\\\nonumber
&=\frac{1}{d} \sum_{(m^{\prime},n^{\prime})\in\Z_d\times\Z_d}\vap(m^{\prime},n^{\prime})\,\overline{F_{s}[f]}(m^{\prime},n^{\prime}) (U(m^{\prime},n^{\prime})\psi)(l)\\\nonumber
&=\frac{1}{d} \sum_{(m^{\prime},n^{\prime})\in\Z_d\times\Z_d}\vap(m^{\prime},n^{\prime})\,\overline{F_{s}[f]}(m^{\prime},n^{\prime}) e^{-\ii\frac{\pi}{d}m^{\prime}n^{\prime}}e^{\ii\frac{2\pi}{d}m^{\prime}l}\psi(l-n^{\prime})\\\nonumber
&=\frac{1}{d} \sum_{(m^{\prime},l^{\prime})\in\Z_d\times\Z_d}\vap(m^{\prime},l-l^{\prime})\,\overline{F_{s}[f]}(m^{\prime},l-l^{\prime}) e^{-\ii\frac{\pi}{d}m^{\prime}(l-l^{\prime}))}e^{\ii\frac{2\pi}{d}m^{\prime}l}\psi(l^{\prime})\\\nonumber
&=\sum_{l^{\prime}\in\Z_d}\left[\frac{1}{d} \sum_{m^{\prime}\in\Z_d}\vap(m^{\prime},l-l^{\prime})\,\overline{F_{s}[f]}(m^{\prime},l-l^{\prime}) e^{-\ii\frac{\pi}{d}m^{\prime}(l-l^{\prime}))}e^{\ii\frac{2\pi}{d}m^{\prime}l}\right]\psi(l^{\prime})\\\nonumber
&=\sum_{l^{\prime}\in\Z_d}\left[\frac{1}{d} \sum_{m^{\prime}\in\Z_d}\vap(m^{\prime},l-l^{\prime})\,\overline{F_{s}[f]}(m^{\prime},l-l^{\prime}) e^{\ii\frac{\pi}{d}m^{\prime}(l+l^{\prime})}\right]\psi(l^{\prime})\\\nonumber
\end{align*}
Finally one gets:
\begin{equation}
\mathcal A_{1}^{\vap}(l, l^{\prime})= \frac{1}{d} \sum_{m\in\Z_d}\vap(m,l-l^{\prime})\,\overline{F_{s}[f]}(m,l-l^{\prime}) e^{\ii\frac{\pi}{d}m(l+l^{\prime})}÷,. 
\end{equation}
\eprf
One easily checks that $f=1$ gives $\overline{F_{s}[f]}(m,l-l^{\prime})=d\delta_{m\,0}\delta_{l\,l^{\prime}}$. Inserting this in $\eqref{kernel_A_f}$ gives:
\begin{equation}
\label{kernel_A_1}
\mathcal A_{1}^{\vap}(l, l^{\prime})
= \sum_{m\in\Z_d} \delta_{m\,0}\delta_{l\,l^{\prime}}
\vap(m,l-l^{\prime})\,e^{\ii\,\frac{\pi}{d}m(l+l^{\prime})}
= \delta_{l l^{\prime}}\vap(0,0)
= \delta_{l\,l^{\prime}}\\
\end{equation}
for $\vap(0,0)=1$.

Covariance in the sense given by Eq. \eqref{covar}  is easily proved in the present case.
\begin{proposition}
The quantization map $f\mapsto A^{\vap}_f$ is covariant with respect to the action of the representation $V$, that is,
\begin{equation}
\label{qmapV}
V\,A^{\vap}_f V^{\dag}= \,A^{\vap}_{\mathcal{V}f}
\end{equation}
where $(\mathcal{V}(s,m_0,n_0)f)(m,n)= f(m-m_0,n-n_0)$ is the induced action  on the phase space.
\end{proposition}



\subsubsection{Quantization of separable functions}
As a first application, we consider the quantization of a separable function:
\begin{equation}
f(k,l)=g(k)\,h(l)\,.
\end{equation}
In this case the function $F$ in \eqref{kernel_A_f} and the ``integral" kernel read as
\begin{equation}
\label{Fgh}
\begin{split}
\overline{F_{s}[f]}(m,l-l^{\prime})&= \frac{1}{d} \sum_{(m^{\prime},n^{\prime})\in\Z_d\times\Z_d}\, g(m^{\prime})h(n^{\prime})e^{\ii \frac{2\pi}{d}(m^{\prime}(l-l^{\prime})-mn^{\prime})}
\\ &= \hat{g}(-(l-l^{\prime}))\hat{h}(m)\,.
\end{split}
\end{equation}
Inserting in $\eqref{kernel_A_f}$ gives:
\begin{equation}
\label{A_f_separab}
\mathcal{A}_{gh}^{\vap}(l-l^{\prime}) = \hat{g}(-(l-l^{\prime})) \frac{1}{d}  \sum_{m\in\Z_d} \hat{h}(m)\vap(m,l-l^{\prime})e^{\ii\,\frac{\pi}{d}m(l+l^{\prime})}\,.
\end{equation}

\subsubsection{Quantization of functions of momentum only}
The restriction  of \eqref{Fgh} and \eqref{A_f_separab} to a function of momentum only, $f(m,n)=g(m)$, yields for the function $F$ and the corresponding integral kernel:
\begin{equation}
\overline{F_{s}[f]}(m,l-l^{\prime})=\hat{g}(-(l-l^{\prime}))\sqrt{d}\,\delta_{m\,0}\,,
\end{equation}
\begin{equation}
\label{kernel_A_{g}}
\begin{split}
\mathcal{A}_{g}^{\vap}(l, l^{\prime})& =  \frac{1}{\sqrt{d}}\sum_{n\in\Z_d} \hat{g}(-(l-l^{\prime}))\delta_{m\,0}\vap(m,l-l^{\prime})e^{-\ii\,\frac{\pi}{d}m(l-l^{\prime})}\\
&= \frac{1}{\sqrt{d}}\hat{g}(-(l-l^{\prime}))\vap(0,l-l^{\prime})\,,
\end{split}
\end{equation}
where $\hat{g}(-k)=\frac{1}{\sqrt{d}}\sum_{m\in\Z_d}g(m)e^{\ii\,\frac{2\pi}{d}mk}$.
There results the following action of the operator $A^{\vap}_g$:
\begin{equation}
\label{Ag}
(A^{\vap}_g\psi)(l)= \frac{1}{\sqrt{d}}\sum_{l^{\prime}}\hat{g}(l^{\prime}-l)\vap(0,l-l^{\prime}) \psi(l^{\prime})\,.
\end{equation}
Let us consider powers of the momentum that $f(m,n)=g(m)=m^L$, $L$ positive integer.
\begin{equation}
\label{ }
\hat{g}(l^{\prime}-l)=\frac{1}{\sqrt{d}}\sum_{k=0}^{d-1} k^{L} e^{\ii\frac{2\pi}{d}k(l^{\prime}-l)}\, . 
\end{equation}
Then 
\begin{align}
\nonumber (A^{\vap}_g\psi)(l)&= \frac{1}{\sqrt{d}}\sum_{l^{\prime}}\hat{g}(l^{\prime}-l)\vap(0,l-l^{\prime}) \psi(l^{\prime})\\\nonumber
&= \frac{1}{d}\sum_{k=0}^{d-1} k^{L}\sum_{l^{\prime}}\vap(0,l-l^{\prime})e^{\ii\frac{2\pi}{d}k(l^{\prime}-l)}\psi(l^{\prime})\\\nonumber
&= \frac{1}{d}\sum_{k=0}^{d-1} k^{L}\sum_{l^{\prime}}\vap(0,l-l^{\prime})e^{\ii\frac{2\pi}{d}(-k)(l-l^{\prime})}\psi(l^{\prime})\\
&= \frac{1}{\sqrt{d}}\sum_{k=0}^{d-1} k^{L}(\vap(0,.)e^{d}_{-k}*\psi)(l)\,. 
\end{align}
In particular:

\begin{itemize}
\item $g(m)=m$
\begin{align}
\label{A_{m}}
\hat{g}(l^{\prime}-l)=\frac{1}{\sqrt{d}}\sum_{k=0}^{d-1} k\,e^{\ii\frac{2\pi}{d}k(l^{\prime}-l)}
\end{align}
\item $g(m)=m^2$
\begin{align}
\hat{g}(l^{\prime}-l)=\frac{1}{\sqrt{d}}\sum_{k=0}^{d-1} k^{2} e^{\ii\frac{2\pi}{d}k(l^{\prime}-l)}
\end{align}
\end{itemize}
\subsection{Quantization of function of  position only}
We now turn to the quantization of a function of the position only, $f(m,n)= h(n)$.
From
\begin{equation}
\hat{g}(l^{\prime}-l)= \sqrt{d}\delta_{l\,l^{\prime}}\, ,
\end{equation}
we obtain for the integral kernel:
\begin{equation}
\label{kernel_A_g}
\mathcal{A}_{h}^{\vap}(l, l^{\prime})
= \sqrt{d}\delta_{l l^{\prime}}\sum_{m\in\Z_d} \hat{h}(m)\, \vap(m,l-l^{\prime})e^{\ii\,\frac{\pi}{d}m(l+l^{\prime})}\,.
\end{equation}
This yields the multiplication operator
\begin{equation}
\label{A_multiply)}
(A^{\vap}_h\psi)(l)= \left(\frac{1}{\sqrt{d}}\sum_{m\in\Z_d} \hat{h}(m)\, \vap(m,0)e^{\ii\,\frac{2\pi}{d}ml}\right)\psi(l)\, ,
\end{equation}
that is,
\begin{align}
\label{A_h(n)}
A^{\vap}_h &= \frac{1}{\sqrt{d}}\sum_{m\in\Z_d} \hat{h}(m)\, \vap(m,0)e^{\ii\,\frac{2\pi}{d}m\,l}\,\\\nonumber.
&= \frac{1}{\sqrt{d}} \sum_{k} h(k)\, [\frac{1}{\sqrt{d}}\sum_{m\in\Z_d}\vap(m,0)e^{\ii\,\frac{2\pi}{d}m\,(l-k)}]\,\\\nonumber.
&= \frac{1}{\sqrt{d}} \sum_{k} h(k)(\mathcal{F}^{-1}\vap_{p_1})(l-k)=\frac{1}{\sqrt{d}}(h*\mathcal{F}^{-1}\vap_{p_1})(l).\\\nonumber.
\end{align}

\subsection{Quantization of a probability distribution}

We consider a probability function distribution $f(p,q)$ on the phase space. Since we are dealing with a discrete space, $f(p, q)$ should satisfy the normalization condition:
\begin{equation}\sum_{q=0}^{d-1} f(p, q) = 1.
\end{equation} This ensures that the total probability over the entire phase space is 1.
The quantized operator $A^{\vap}_f$ is density operator if and only if, for any nonzero vector $\psi\in L^2(\Z_d)$,
\begin{align*}
&\langle\,\psi|A^{\vap}_f\psi\rangle
= \frac{1}{d}\sum_{(m,n)\in\Z_d\times\Z_d} f(m,n)\, \langle\psi|\sfMv(m,n)\psi\rangle\geq 0\, . 
\end{align*}
A simple example is the case where $\vap$ where is a coherent state weight: that is, $\vap(m,n)=\langle U(m,n)\phi|\phi\rangle$. In this case:
\begin{equation}
\langle\psi|\sfMv(m,n)\psi\rangle=|\langle\,U(m,n)\phi|\psi\rangle|^2
\end{equation}
which is the Husimi distribution.

NB: One can be more general by replacing $\sfMv(m,n)$ by the $\rho(m,n)=U(m,n)\rho_0\,U(m,n)^{\dagger}$, the transported by $U(m,n)$, of any density operator $\rho_0$ on $L^2(\Z_d)$.


\section{Semi-classical portrait}
\label{portrait}
Given a function $\vap(m,n)$ on the phase space $\Gamma$, normalised at $\vap(0,0)=1$, and yielding a positive unit trace operator, i.e. a density operator, $\sfMv $, the quantum phase space portrait of an operator $A$ on $L^2(\Gamma,\ud\gamma)$ reads as:
\begin{equation}
\label{semclA}
\check{A}(m,n):=  \mathrm{Tr}\left(A\,U(m,n)\,\sfMv U^{\dag}(m,n )\right) = \mathrm{Tr}\left(A\,\sfMv (m,n)\right)\, .
\end{equation}
The most interesting aspect of this notion in terms of probabilistic interpretation holds when the operator $A$ is precisely the integral quantized version $A^\vap_f$ of a classical $f(m,n)$ with the same function $\vap$ (actually we could define the transform with 2 different ones, one for the ``analysis'' and the other for the ``reconstruction''). Then, with the  use of the composition rule \eqref{UUp} we compute the transform:
\begin{align*}
\label{lowsymbv}
f( m,n )&\mapsto \check f(m,n) \equiv \check{A}_f^{\vap}( m,n )=\mathrm{Tr}\left(A^{\vap}_{f}\,\sfMv (m,n)\right)\\\nonumber
&=\frac{1}{d}\sum_{(m^{\prime},n^{\prime})\in\Z_d\times\Z_d}\,f(m^{\prime},n^{\prime} )\mathrm{Tr}\left(U(m^{\prime},n^{\prime})\sfMv U^{\dag}(m^{\prime},n^{\prime})\,U(m,n)\sfMv U^{\dag}(m,n)\right)\\\nonumber
&=\frac{1}{d}\sum_{m^{\prime},n^{\prime}\in\Z_d\times\Z_d} f(m^{\prime},n^{\prime} )\mathrm{Tr}\left(\sfMv \,U(m-m^{\prime},n-n^{\prime})\sfMv U(-(m-m^{\prime}),-(n-n^{\prime}))\right)\\\nonumber
&=\frac{1}{d}\sum_{(m^{\prime},n^{\prime})\in\Z_d\times\Z_d} f(m-m^{\prime},n-n^{\prime} )\mathrm{Tr}\left(\sfMv \,U(m^{\prime},n^{\prime})\sfMv U^{\dag}(m^{\prime},n^{\prime})\right)\\\nonumber
&=\sum_{(m^{\prime},n^{\prime})\in\Z_d\times\Z_d} f(m-m^{\prime},n-n^{\prime} )\mathrm{Tr}\left(\sfMv\,\sfMv (m^{\prime},n^{\prime})\right)\,.
\end{align*}
Finally we get:
\begin{proposition}
The semi-classical portrait of the operator $A^{\vap}_{f}$ with respect to the weight $\vap$ is given by:
\begin{equation}
\label{lowsymbv}
\check f(m,n)
=\frac{1}{d}\sum_{(m^{\prime},n^{\prime})\in\Z_d\times\Z_d}\, f(m-m^{\prime},n-n^{\prime} )\mathrm{Tr}\left(\sfMv (m^{\prime},n^{\prime})\,\sfMv\right)\,.
\end{equation}
\end{proposition}
This expression, which can be viewed as a convolution on the phase space has the meaning of an averaging of the classical $f$. The function
\begin{align}
\label{distvap}
(m,n) &\mapsto \mathrm{Tr}\left(\sfMv(m,n)\sfMv\right)\, ,
\end{align}
is a true probability distribution on $\Gamma$, i.e. is positive and with integral on $\Gamma$ equal to 1,    from the resolution of the identity and the fact  that $\sfMv$ is chosen as a density operator. Expression \eqref{distvap} is actually a kind of Husimi function.
The  expression of $\mathrm{Tr}\left(\sfMv(m,n )\sfMv\right)$  is easily derived and reads as
\begin{equation}
\label{trmommom}
\begin{split}
\mathrm{Tr}\left(\sfMv(m,n )\sfMv\right)
&=\frac{1}{d}\sum_{k\in\Z_d}\langle\,e_{k}\vert\left(\,\sfMv (m,n)\sfMv \right)\,e_{k}\rangle\\
&=\frac{1}{d^2}\sum_{(m^\prime,n^{\prime})\in\Z_d\times\Z_d}\vert\vap(m^{\prime},n^{\prime})\vert^{2}e^{\ii \frac{2\pi}{d}\,(mn^{\prime}-m^{\prime}n)}\,.
\end{split}
\end{equation}
Summing this expression on $\Gamma$, we get $1$, which means that $\check 1= 1$, as expected.

\section{Quantization with various weights}
\label{weight}

\subsection{Weight related to  coherent states for \texorpdfstring{$H(\Z_d)$}{H(Zd)}}

The weight $\vap_\phi$ corresponding to the projection operator $\vert\phi\rangle\langle\phi\vert$ of a square integrable function $\phi$, with unit norm,  on $\Z_d$ is  found through  the trace formula given by Proposition \ref{tretrw}:
\begin{equation}
\label{weight_cs}
\vap_{\phi}(m,n)= \mathrm{Tr}[U^{\dag}(m,n)\vert\phi\rangle\langle\phi\vert]= \langle U(m,n)\phi\vert\phi\rangle
\end{equation}
In this case, from $\vap_\phi=\langle U(m,n)\phi\vert\phi\rangle$, and from the Fourier expansion of  $\phi \in L^2(\Z_d)$,  
\begin{equation}
\label{fourphi}
\phi = \sum_{k} \hat{\phi}(k) e^{d}_k\, , \quad \hat{\phi}(k)= \lg e^{d}_k|\phi\rg=  \sum_{n\in\Z_d}\frac{ e^{-\ii \frac{2\pi}{d}n k}}{\sqrt{d}}\, \phi(n)\, ,
\end{equation}
we have for $\Omega_{\phi}(n)= \vap_\phi(0,n)$:
\begin{align}
&\Omega_{\phi}(n)= \langle U(0,n)\phi\vert\phi\rangle=\sum_{n\in\Z_d}e^{\ii\,\frac{2\pi}{d}mn}\vert\hat{\phi}(n)\vert^2=\left\lg e^{\ii\,m\cdot}\right\rg_{\vert\phi\vert^2}\,,\\
&\Omega_{\phi}^{\prime}(0)=\ii\sum_{m\in\Z_d}\,m\,\vert\hat{\phi}(m)\vert^2=\ii \lg m\rg_{\vert\phi\vert^2}\,, \\
&\Omega_{\phi}^{\prime\prime}(0))=-\sum_{m\in\Z_d}\,m^2\vert\hat{\phi}(m)\vert^2 = -\lg m^2\rg_{\vert\phi\vert^2}\,,
\end{align}
where $\lg\cdot\rg_{\vert\phi\vert^2}$ means the average of the random variable ``$\cdot$'' with respect to the discrete probability distribution  $\Z\ni m\mapsto \vert\hat{\phi}(m)\vert^2$ ($\phi$ has unit norm).
We then derive  the following quantizations of the  angular momentum, of its square, and a function of the angle only.
\begin{itemize}
\item $g(m)=m$, and from $\eqref{Ag}$, we get:
\begin{equation}
(A^{\vap}_g\psi)(l)= \frac{1}{\sqrt{d}}\sum_{k=0}^{d-1} k\,(\Omega_{\phi}(m)e^{d}_{-k}*\psi)(l)\,.
\end{equation}
\item $g(m)=m^2$ and from $\eqref{Ag}$, we get:
\begin{equation}
(A^{\vap}_g\psi)(l)= \frac{1}{\sqrt{d}}\sum_{k=0}^{d-1} k^{2}(\Omega_{\phi}(m)e^{d}_{-k}*\psi)(l)\, .
\end{equation}
\item $h(n)$ and from $\eqref{weight_cs}$ we get the multiplication operator:
\begin{equation}
\label{A_f(n)}
\begin{split}
\left(A^{\vap_{\phi}}_h\psi\right)(l)&= \left[\sum_{m\in\Z} \hat{h}(m)\, \langle\,U(m,0)\phi\vert\phi\rangle\,e^{\ii\,ml}\right]\psi(l)\\
&= \left[ \sum_{m\in\Z} \hat{h}(m)\,\lg e^{-\ii ml}\rg_{\vert\phi\vert^2}\,e^{\ii\,m\gamma}\right]\psi(l)\,.
\end{split}
\end{equation}
\end{itemize}

\subsection{Weight related to the parity operator}

The weight $\vap_{\sfP}$ corresponding to the parity operator $\sfP\psi(l)= \psi(-l)$ on a function $\psi$ in $L^{2}(\Z_d)$ is simply:
\begin{equation}
\label{varpi_parity}
\vap_{\sfP}(m,n)= 1\,.
\end{equation}
and 
\begin{equation}
(\sfMvP\psi)(l)= \psi(-l), \,.
\end{equation}
NB: In this subsection, we consider the case where $d$ is {\it odd}
\bprf
First, we have:
\begin{align}
&(\sfMvP\psi)(l)= \frac{1}{d}\sum_{(m,n)\in\Z_d\times\Z_d}(U(m,n)\psi)(l)\\\nonumber 
&= \frac{1}{d}\sum_{(m,n)\in\Z_d\times\Z_d}e^{-\ii\frac{\pi}{d}mn}
e^{\ii\frac{2\pi}{d}ml}\psi(l-n),\\\nonumber
&=\frac{1}{d}\sum_{n\in\Z_d}\sum_{m\in\Z_d}e^{\ii\frac{2\pi}{d}m(l-\frac{n}{2})}\psi(l-n)\\\nonumber
&=\frac{1}{d}\sum_{n^{\prime}\in\Z_d}\sum_{m\in\Z_d}e^{\ii\frac{2\pi}{d}m(l-n^{\prime})}\psi(l-2n^{\prime})\\\nonumber
&=\frac{1}{d}\sum_{n^{\prime}\in\Z_d} d\,\delta_{l,n^{\prime}}\psi(l-2n^{\prime})=\psi(-l)\,. 
\end{align}
It is to be noted that $\frac{1}{2}\equiv \frac{d+1}{2}$ is the inverse of $2$, because d is odd. Then, 

\begin{align*}
&\,\mathrm{Tr}[U^{\dag}(m,n)\sfP]= \mathrm{Tr}[U(m,n)\sfP]\\\nonumber
&=\sum_{l\in\Z}\langle\,e_l\vert\,U(m,n)^{\dag}\sfP\,e_l\rangle=\sum_{l\in\Z}\langle\,U(m,n)\,e_l\vert\,\sfP\,e_l\rangle\\\nonumber
&=\frac{1}{d}\sum_{(l,k)\in\Z_d\times\Z_d} e^{\ii\frac{\pi}{d}mn}\,e^{-\ii\,\frac{2\pi}{d}mk}\,e^{-\ii\,\frac{2\pi}{d}l(k-n)} e^{\ii\,\frac{2\pi}{d}l(-k)}\\\nonumber
&=\frac{1}{d}\sum_{(l,k)\in\Z_d\times\Z_d} e^{\ii\frac{\pi}{d}mn}\,e^{-\ii\,\frac{2\pi}{d}mk}\,e^{-\ii\,\frac{2\pi}{d}l(2\,k-n)} \\\nonumber
&=\frac{1}{d}\sum_{l\in\Z_d}e^{-\ii\,\frac{\pi}{d}\,m(2l-n)}d\delta_{2l-n,0}
=1\,.
\end{align*}
\eprf
Defining a Wigner-like  distribution \cite{mukunda1, mukunda2, grossmann} of $\psi$ on the phase $\Z_d\times\,\Z_d$ as
\begin{equation}
W_\psi(m,n) =   \frac{1}{d}\sum_{l\in\Z_d}\,e^{\ii\frac{4\pi}{d}ml} \overline{\psi(n+l)}\psi(n-l)\, , 
\end{equation}
 then, in the case where $d$ is {\it odd}, we have:
\begin{equation}
W_\psi(m,n) =   \frac{1}{d}\sum_{l\in\Z_d}\,e^{\ii\frac{2\pi}{d}ml} \overline{\psi(n+\frac{l}{2})}\psi(n-\frac{l}{2}).
\end{equation}

Let us see how it is linked to the parity operator $\sfP$ and its corresponding uniform weight $\vap_{\sfP}(m,n)= 1$.
\begin{proposition}
The Wigner-like  distribution of $\psi$ on the phase $\Z_d\times\,\Z_d$ is the mean value $\times 1/d$ of the Weyl-Heisenberg transport of the parity operator in the state $\psi$ :
\begin{equation}
\label{WeylParity}
W_\psi(m,n) = \frac{1}{d} \langle\psi| \sfP(m,n)\psi\rangle\, , \quad \mbox{with}\quad  \sfP(m,n) = U(m,n)\sfP U(m,n)^{\dag}\,.
\end{equation}
\end{proposition}
\bprf
\begin{align*}
W_\psi(m,n) &= \frac{1}{d} \langle\psi\vert\,\sfP (m,n)\psi\rangle= \frac{1}{2\pi} \langle\psi\vert\,{U}(m,n)\, \sfP \, {U}^{\dag}(m,n)\psi\rangle\\\nonumber
&=  \frac{1}{d}\langle\,{U}^{\dag}(m,n))\psi\vert\, \sfP \, {U}^{\dag}(m,n)\psi\rangle\\\nonumber
&=  \frac{1}{d}\sum_{l\in\Z_d}\,\overline{{U}^{\dag}(m,n))\psi(l)}\, \sfP \, {U}^{\dag}(m,n)\psi(l)\\\nonumber
&=  \frac{1}{d}\sum_{l\in\Z_d}\,\overline{{U}^{\dag}(m,n))\psi(l)} \, {U}^{\dag}(m,n)\psi(l^{\prime}), \, l^{\prime}=-l, \\\nonumber
&=  \frac{1}{d}\sum_{l\in\Z_d}\,e^{\ii\frac{\pi}{N}mn} e^{\ii\frac{2\pi}{d}ml} \overline{\psi(l+n)} \, e^{-\ii\frac{\pi}{N}mn} e^{-\ii\frac{2\pi}{d}ml^{\prime}} \psi(l^{\prime}+n), \, l^{\prime}=-l, \\\nonumber
&=  \frac{1}{d}\sum_{l\in\Z_d}\,e^{\ii\frac{\pi}{d}mn} e^{\ii\frac{2\pi}{d}ml} \overline{\psi(l+n)}  \, e^{-\ii\frac{\pi}{d}mn} e^{\ii\frac{2\pi}{d}ml} \psi(-l+n), \\\nonumber
 &=  \frac{1}{d}\sum_{l\in\Z_d}\,e^{\ii\frac{4\pi}{d}ml} \overline{\psi(n+l)}\psi(n-l).\, \\\nonumber
\end{align*}
For $d$ {\it odd}, $2l=l^{\prime}$ gives $l=\frac{d+1}{2} l^{\prime}$, we get:
\begin{align}
    W(m,n) &=  \frac{1}{d}\sum_{l^{\prime}\in\Z_d}\,e^{\ii\frac{2\pi}{d}ml^{\prime}} \overline{\psi\left(n+\frac{d+1}{2} l^{\prime}\right)}\psi\left(n-\frac{d+1}{2} l^{\prime}\right) \\\nonumber
    &=  \frac{1}{d}\sum_{l^{\prime}\in\Z_d}\,e^{\ii\frac{2\pi}{d}ml^{\prime}} \overline{\psi\left(n+\frac{1}{2} l^{\prime}\right)}\psi\left(n-\frac{1}{2} l^{\prime}\right)\, . 
\end{align}
\eprf
Given a state $\psi$, one can also show that its Wigner distribution can be retrieved  through a symplectic Fourier transform type of  its reproducing kernel. For $ \vap=\vap_{\sfP}=1$, one has

\begin{align*}
&A^{1}_f
= \frac{1}{d}\sum_{(m,n)\in\Z_d\Z_d} f(m,n )\, \sfMv(m,n) \\
&=\frac{1}{d^2} \sum_{(m^{\prime},n^{\prime})\in\Z_d\times\Z_d} \sum_{(m,n)\in\Z_d\times\Z_d} f(m,n)\, U(m,n)U(m^{\prime},n^{\prime})U^{\dag}(m,n)\\\nonumber
&=\frac{1}{d^2} \sum_{(m^{\prime},n^{\prime})\in\Z_d\times\Z_d} \sum_{(m,n)\in\Z_d\times\Z_d} f(m,n )\, e^{\ii\frac{2\pi}{d}(mn^{\prime}-m^{\prime}\,n)}U(m^{\prime},n^{\prime})\\\nonumber
\end{align*}

Hence we get for the quantizations of elementary functions: 
\begin{itemize}
\item Function of the moment only: $f(m,n)=g(m)$.  The kernel is given by:
\begin{align}
\label{kernel_A_{g}_wigner}
\mathcal{A}_{g}^{1}(l,l^{\prime})  =\frac{1}{\sqrt{d}}\hat{g}(l^{\prime}-l)\,,
\end{align}
\end{itemize}
and this gives:
\begin{equation}
\label{A_{g}_wigner}
(A_{g}^{1}\psi)(l)=(FgF^{-1}\psi)(l)\,, 
\end{equation}
i.e., $ A_{g}^{1}=FgF^{-1}$.
\begin{itemize}
\item Function of position only.  The kernel reads:
\end{itemize}
\begin{equation}
\mathcal{A}_{h}^{\vap}(l, l^{\prime})
= \frac{1}{\sqrt{d}}\delta_{l l^{\prime}}\sum_{m\in\Z_d} \hat{h}(m)\,e^{\ii\,\frac{2\pi}{{d}}ml^{\prime}}=\delta_{l,l^{\prime}}h(l)\,,
\end{equation}
which means that $A_h^{1}$ is multiplication operator by $h$,
\begin{equation}
(A_h^{1}\psi)(l)=h(l)\psi(l)\,. 
\end{equation}
In this case: the quantized position on the discrete circle, i.e., the discrete angle, is $h(l) = \frac{2\pi l }{d}$ or its powers, $h(l) = l^L, L\in\N^{*}$, are discontinuous. One may also periodize $h(l) = h(l \pm d)$. However, one gets the same discontinuity in the operator as in the original function. There is no regularization.

An alternative would be to consider position operator through the multiplication by the smooth Fourier exponential 
\begin{align}
    h(l) = e^{\ii\,\frac{2\pi l}{d}}.
\end{align}

\section{Conclusion}
\label{conclu}

In this paper, we have established a covariant  integral quantization for  systems whose phase space is $\Z_d\times\,\Z_d$, i.e., for systems moving on the discrete circle.  

The symmetry  group of this phase space is the discrete  version of the Weyl-Heisenberg group, namely  the central extension of the abelian group $\Z_d\times\,\Z_d$,  and the phase space can be viewed as the left coset of the group with its center. The existence of a non-trivial unitary irreducible representation  of this group on the phase phase, as acting on $L^2(\Z_d)$, allowed us to derive interesting results. First, we have established the concomitant resolution  of the identity and subsequent properties such the Gabor transform on the discrete circle and its inversion, the reproducing kernel and the fact that any square integrable function on the circle is  fiducial vector. The decomposition of the identity allows to define an integral operator $\mathsf{M}^{\vap}$ from weight function $\vap$ defined on the phase space.


There are noticeable results related to the quantization of a point $(m,n)$ in the phase space according to two standard choices of the weight.

\begin{itemize}
\item With the parity weight, the quantization of the momentum is transported the momentum by the Fourier operator F, that is
\begin{equation}
m\mapsto\hat{m}= FmF^{-1}\,, (A_{m}^{1}\psi)(l)= (FmF^{-1}\psi)(l)
\end{equation}
while the quantization of the discrete angle yields the multiplication operator by the $h(n)=\frac{2\pi\,n}{d}$.
\begin{equation}
\frac{2\pi\,n}{d}\mapsto A_{h(n)}^{1}\, , \quad (A_{h(n)}^{1}\psi)(l)=\frac{2\pi\,l}{d}\psi(l)\,.
\end{equation}
This is of course not acceptable. Alternatively one can quantize the periodized  angle function  $(\mathrm{Per}\,\mathsf{A})(\theta), \theta=\frac{2\pi\,n}{d}$. One finds the multiplication operator defined by the same discontinuous $(\mathrm{Per}\,\mathsf{A})(\gamma), \gamma= \frac{2\pi\,n}{d}$. There is no regularization.
\\
\item With the coherent state weight one obtains the quantization,
\begin{equation}
m\mapsto\hat{m}= 
(A^{\vap}_g\psi)(l)= \frac{1}{\sqrt{d}}\sum_{k=0}^{d-1} k\,(\Omega_{\phi}(m)e^{d}_{-k}*\psi)(l)\\\nonumber
\end{equation}
whereas the quantization of the periodized function $\theta=\frac{2\pi\,n}{d}$ leads to its smooth regularization,  
\begin{equation}
(\mathrm{Per}\,\mathsf{A})(n)\mapsto\left[ \sum_{m\in\Z} \hat{h}(m)\,\lg e^{-\ii \frac{2\pi\,m}{d}l}\rg_{\vert\phi\vert^2}\,e^{\ii\,\frac{2\pi\,m}{d}\,l}\right]\psi(l)\,.
\end{equation}

\end{itemize}

In a follow up works, we  plan to extend the results of this work in three directions:
\begin{itemize}
\item Feature detection in signals defined on $Z_d$ using appropriate quantum operators derived from the above quantization formalism. In particular to quantum period finding\cite{Howard}.
\item Investigate the usefulness of this formalism in quantum computation, quantum information, and cryptography \cite{Wang}. This includes exploring problems such as period finding, entanglement, and the design of symmetric informationally complete bases or fiducial vectors\cite{Miquel, Dittel, Lawrence, Maurice, Appleby}. 
\item Covariant integral quantization of an affine discrete phase space, $\Z_{d}\times\Z_d^{*}$, the discrete affine group where $d$ is prime. This phase space has a group structure of translations and dilations on $\Z_{d}$ with d prime. The related wavelet transform was studied by Flornes, Grossmann, Holschneider, Torr\'esani \cite{Flornes, grossmann}.
\end{itemize}

\section*{Appendix}
In this appendix, we give an example that illustrates the importance of phase space representations for signals. In this case, the discrete Weyl-Heisenberg Gabor transform Eq. \eqref{pspphi} of a signal with respect to the Von Mises fiducial vector as seen in equation Eq. \eqref{VonMises}. We consider two periodic signals, one with a period of length 6 and one with a period of length 15. Signal 1 is given as $[\dotsb, 2,4,4,3, 3, 5,\dotsb]$ and signal 2 is given as $[\dotsb, 0, 0, 0, 3, 3, -2, 1, 1, 1, 4, -1, -1, 2, 2, 2, \dotsb]$. Their sum will be a new signal with a period of 10, given by $[\dotsb, 2, 4, 4, 6, 6, 3, 3, 5, 5, 7, \dotsb]$. Here we take the sum of signals 1 and 2, both with a length of 60. This gives us signal 3 of length 60, with period 10. These examples are discussed in \cite{Restrepo}. 
\begin{figure}[H]
    \centering
    \includegraphics[angle=0,width=6cm]{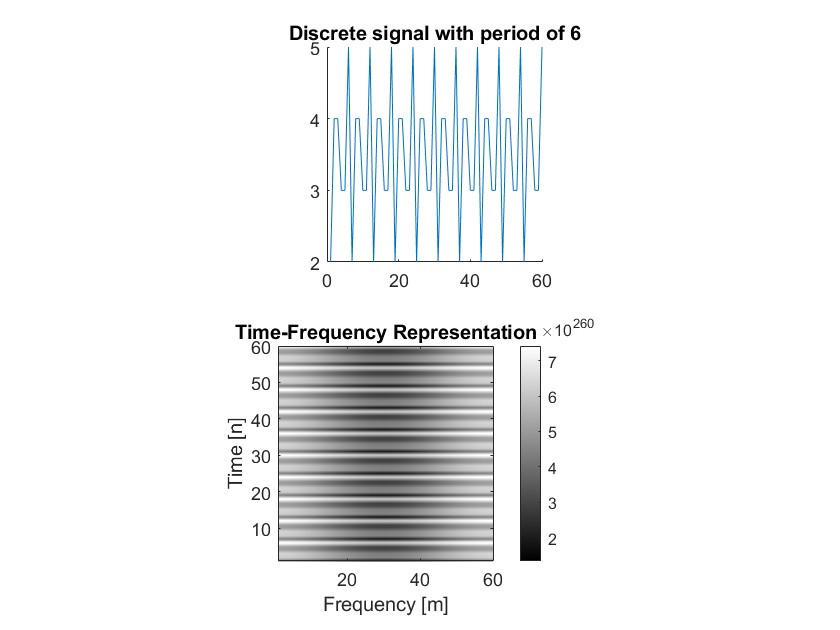}
    \includegraphics[angle=0,width=6cm]{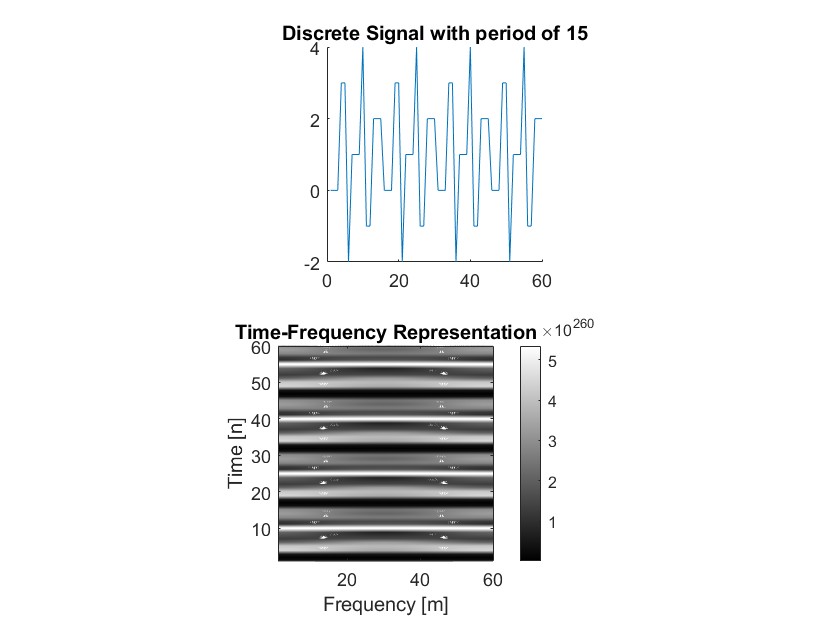}
    \includegraphics[angle=0,width=8cm]{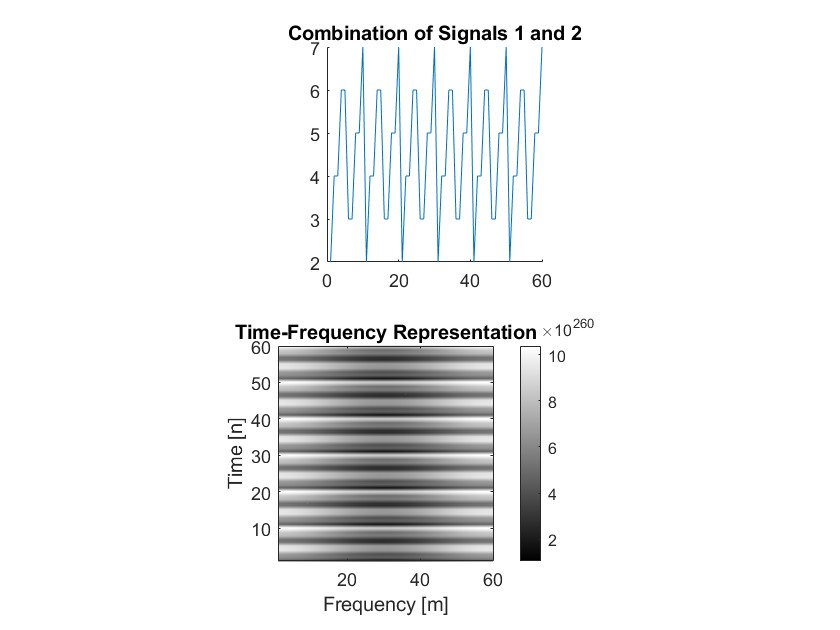}
    \caption{The top left \& top right plots show two discrete signals with different periods and their following time-frequency representations. Below shows the combination of the two signals, yielding a unique period from either of the original two signals. The Gabor transformation of signal 3 shows that its period, found by the number of striations in the plot, is different than either signal 1 or signal 2. All 3 time-frequency representations use the Von Mises filter \eqref{VonMises} scaled at $\lambda = 400$.}
    \label{gaborSignals}
\end{figure}

\section*{Acknowledgements}
R. Murenzi, A. Zlotak, and J. P. Gazeau are grateful for the funding provided by the Physics Department of the Worcester Polytechnic Institute. In addition, J.-P. Gazeau is thankful for the financial support that allowed him to travel to Worcester, MA to further the collaboration on this work.

\end{document}